# Trends in sea-ice variability on the way to an ice-free Arctic

S. Bathiany[1], B. van der Bolt[1], M. S. Williamson[2], T. M. Lenton[2], M. Scheffer[1], E. van Nes[1], D. Notz[3]

[1]Department of Aquatic Ecology and Water Quality Management, Wageningen University, NL-6700 AA Wageningen, The Netherlands
[2]College of Life and Environmental Sciences, University of Exeter, United Kingdom
[3]Max-Planck-Institute for Meteorology, Bundesstrasse 53, 20146 Hamburg, Germany

*Correspondence to*: S. Bathiany (sebastian.bathiany@wur.nl)

**Abstract.** It has been widely debated whether Arctic sea-ice loss can reach a tipping point, beyond which a large sea-ice area disappears abruptly. An important argument for this scenario is the destabilising role of the ice-albedo feedback. The theory of dynamical systems predicts a 'slowing down' when a system destabilises towards a tipping point. In simple stochastic systems this can result in increasing variance and autocorrelation, potentially yielding an early warning of an abrupt change. Here we aim to establish whether the loss of Arctic sea ice would follow these conceptual predictions, and which trends in sea ice variability can be expected from pre-industrial conditions toward an Arctic that is ice-free during the whole year.

To this end, we apply a model hierarchy consisting of two box models and one comprehensive Earth system model. We find a consistent and robust decrease of the ice volume's annual relaxation time before summer ice is lost because thinner ice can adjust more quickly to perturbations. Thereafter, the relaxation time increases, mainly because the system becomes dominated by the ocean water's large heat capacity when the ice-free season becomes longer. Both trends carry over to the autocorrelation of sea ice thickness in time series. Also accounting for the geometric effect of increasing open-water formation efficiency for thinning ice, we obtain an increasing variability in sea-ice area fraction, but a decreasing variability in sea-ice thickness.

These changes are robust to the nature and origin of climate variability in the models and hardly depend on the balance of feedbacks. Therefore, characteristic trends can be expected in the future. As these trends are not specific to the existence of abrupt ice loss, the prospects for early warnings seem very limited. This result also has implications for statistical indicators in other systems whose effective 'mass' changes over time, affecting the trend of their relaxation time. However, the robust relation between state and variability would allow an estimate of sea-ice variability from only short observations. This could help one to estimate the likelihood and persistence of extreme events in the future.

# 1 Introduction

Given the rapid loss of Arctic ice area in observations and the potential involvement of large positive feedbacks, the question arises whether Arctic sea ice loss could reach a tipping point at which it accelerates substantially. In the extreme case, this point would correspond to a bifurcation point, a point of no return where sea ice is suddenly and irreversibly lost (Lindsay and Zhang, 2005). It has been suggested that statistical indicators can be used to infer such a destabilisation before the tipping point (Wiesenfeld and McNamara, 1986; Kleinen et al., 2003), making them potential "early warnings" of abrupt change (Scheffer et al., 2009). This idea applies to dynamical systems close to a stable fixed point that slowly destabilises over time. As the forces that restore a disturbed system towards the equilibrium become weaker, the return to equilibrium becomes slower, leading to an increasing relaxation time. In the presence of small perturbations in the form of a stochastic term added on the dynamic equation, this 'slowing down' must cause an increase in autocorrelation and variance when approaching the tipping point. In principle, this concept also applies to systems whose solution is not constant but periodic in time (Wiesenfeld and McNamara, 1986): By recording the state of a system at the same point in time during every period, a periodical solution can be transformed to a fixed point. However, the occurrence of statistical stability indicators relies on several assumptions such as the approximation of the system as one-dimensional (Bathiany et al., 2013), the assumption that the variability of the system results from small white noise external to the system (Dakos et al., 2012b), and that the system is close to its equilibrium solution. None of these assumptions is truly justified in the context of anthropogenic climate change. Moreover, even in cases when all assumptions hold, it is often not clear in practice how close a system needs to be to a bifurcation point for the theory to apply, and how slowly the destabilisation needs to occur to allow a significant detection of trends in variance or autocorrelation. In order to apply the early warning concept to complex systems like climate models, it therefore has to be established whether the concept is applicable. In this study, we explore the applicability of statistical stability indicators to the problem of Arctic sea-ice loss by analysing simulations from models of very different complexity.

It follows already from previous studies that the total hemispheric ice area is not a suitable property to infer sea ice stability. First, the distribution of continents determines where sea-ice can occur and thus determines the variability of total sea-ice area (Goosse et al., 2009; Eisenman, 2010): While sea ice area in the Arctic ocean is free to fluctuate, further south it is limited to the North Atlantic and North Pacific. The rest of the area is covered by continents which therefore 'mute' the variability of total sea ice area (Eisenman, 2010). Second, when the latitude of the sea ice edge approaches the pole, there is less and less total area available in the (idealised) ice-covered circle (Goosse et al., 2009). Third, it has been noted that when sea ice becomes very thin, its open-water formation efficiency increases, meaning that small fluctuations in volume can lead to large fluctuations in area coverage (Holland et al., 2006; Goose et al., 2009; Notz, 2009). As all these effects result from geometrical constraints or conditions of the problem, they do not reflect the stability of the system in terms of its dynamical response to perturbations. We therefore focus on ice thickness in most of this study.

Previous studies mainly relied on very idealised sea-ice models. Using a simple box model, Merrifield et al. (2008) find increasing variance and an increasing relaxation time before an abrupt loss of summer sea ice in their model, apparently corroborating the classical concept of early warnings. However, the seasonal cycle is only parameterised crudely in their model, lumping processes of melting and freezing together in one equation. When resolving a continuous seasonal cycle, Eisenman and Wettlaufer (2009) obtain a stable and gradual loss of summer ice but a bifurcation for the subsequent loss of seasonal ice. In a slightly simplified version of this model, Moon and Wettlaufer (2011, 2013) and Eisenman (2012) showed a relatively complex evolution of the system's timescale over a range of long-wave forcing with decreasing and increasing regimes due to a continuously changing balance of feedbacks. The most important positive feedback in this context is the sea ice-albedo feedback: Due to the ocean's low albedo compared to sea ice, ice loss and decreased surface albedo enhance each other. The most important negative feedback is the growth-thickness feedback (Bitz and Roe, 2004): The thinner the ice becomes, the faster it can regrow due to an increased heat flux through the ice. Moreover, the relatively large timescale of warming or cooling of the ocean's mixed layer becomes important once sea ice is not present during a substantial part of the year. Using an aquaplanet model that is a spatially explicit version of the model by Eisenman and Wettlaufer (2009), Wagner and Eisenman therefore find that the mixed-layer effect can raise false alarms of abrupt ice loss.

So far, it has not been investigated how these factors affect the prospects for early warning signals, especially in more complex, spatially explicit models and the real world. Here we use a hierarchy of models to investigate this question. In particular, we apply two column models that both show abrupt Arctic winter sea-ice loss, but due to different mechanisms. We also modify the models to obtain a gradual sea-ice loss by disabling the ice-albedo feedback. We show that statistical indicators are not specific to these cases and that comprehensive climate models behave in a similar way as these simpler models.

## 2 Models and Methodology

### 2.1 Models

We apply different models in our study that all include a continuous annual cycle, the ice-albedo feedback and the growth-thickness feedback, and that are of very different complexity:

1. the box model by Eisenman (2012) with default parameters, here referred to as E12. The model consists of a simple energy balance of the ocean's mixed layer and describes the evolution of only one state variable, the enthalpy E. In the presence of sea ice, E is negative and proportional to the ice thickness, while during ice-free conditions, E is positive and proportional to the mixed-layer temperature. The model equations are taken from Eisenman (2012) and reproduced in Appendix A. The effect of $CO_2$ is represented implicitly in the surface net longwave balance $L_m$, which is our control parameter for this model. The model yields one stable solution with a perennial ice cover for present-day conditions, $L_m=1.25$ (as the model is non-dimensional, E and $L_m$ have no units). With decreasing $L_m$, the ice becomes thinner and the transition to a seasonal ice cover is gradual (Fig. 1a). In contrast, at $L_m \approx 0.925$, the remaining winter ice disappears abruptly due to a bifurcation in the system. Beyond this bifurcation point the only remaining stable cycle is ice-free during the whole year.

2. the box model by Eisenman (2007), referred to as E07 (see Appendix B). Like E12, it solves the energy balance of the mixed layer, taking solar radiation and atmospheric composition as boundary conditions. Its main difference to E12 is that the model includes more dynamic variables, most importantly an ice-area fraction which is calculated with the parameterization by Hibler (1979). Atmospheric $CO_2$ is prescribed and given as a factor multiplied to the present-day concentration. In similarity to E12, the model shows a gradual loss of summer sea ice, but an abrupt loss of winter sea ice under warming (Fig. 1b, c). The abrupt winter ice loss mainly results from the fact that the large ice area that forms each winter does not form anymore when the ocean no longer cools to the freezing temperature (Bathiany et al., in review). In contrast to a bifurcation that results from a positive feedback, this abrupt change at a threshold is reversible. As the ice-albedo feedback is active also in E07, it produces a regime with multiple solutions around the critical $CO_2$ value. However, this regime is extremely small and thus not practically relevant.

3. the comprehensive Earth system model of the Max-Planck-Institute for Meteorology (MPI-ESM) (Giorghetta et al., 2013). In comparison to the two box models, the Earth system model MPI-ESM is much more complex. As a spatially explicit, comprehensive model, it covers all processes considered relevant for the evolution of sea ice, including mechanical processes and horizontal transport. Despite this huge process complexity, the description of the sea-ice thickness distribution is relatively simple (Notz et al., 2013): Similarly to E07, only one thickness class is used and the ice area is calculated using the parameterisation by Hibler (1979).

4. We also analyse seven additional comprehensive models from the Coupled Model Intercomparison Project 5 (CMIP5), using simulations of the historical period and the RCP8.5 scenario. The latter is the CMIP5 scenario with the largest warming, reaching a radiative forcing of approximately 8.5 $Wm^{-2}$ in the year 2100. The models are all the available models that lose their Arctic winter sea ice in RCP8.5. The level of complexity in these models is comparable to MPI-ESM, but some of them explicitly resolve several ice-thickness classes.

## 2.2 Methods

To investigate how the relaxation time, variance and autocorrelation in the models vary with $CO_2$, we perform three types of experiments that are reported in Sect. 3.1, 3.2 and 3.3 respectively, and whose technical details we address in Appendix A and B:

1. For the two column models E12 and E07 we perform numerical perturbation experiments where we run each model to its equilibrium annual cycle, then suddenly perturb it away from this reference solution by some small amount $x_0$, and measure the rate of return towards the reference solution. It follows from a linearisation of the system that the anomaly x decays exponentially over time t:

$$x = x_0 e^{-(t/\tau)} \qquad \text{(Eq. 1)}$$

In systems with a time-independent solution, the relaxation time scale $\tau$ can essentially be obtained from only one specific state x at a time t by rearranging Eq. (1):

$$\tau = -\frac{t}{\log\left(\left|\frac{x}{x_0}\right|\right)} \qquad \text{(Eq. 2)}$$

Due to the permanent change in the balance of feedbacks during the annual cycle, the return rate in the models we use depends on the time of year. To obtain a good estimate of the anomaly decay from year to year, we store t and x on December 21st in each year (the result is not sensitive to the choice of the date). We obtain the relaxation time τ from a linear regression of these annually resolved time series by regressing the numerator versus the denominator of Eq. (2). For details of this approach when applied to the two column models, see Appendix A and B.

2. For both column models, we perform stationary simulations and calculate the state variables' statistics. For each of 50 different $CO_2$ levels we simulate 100,000 years under constant conditions. We then compute seasonal means for winter sea ice (averages over March to May), summer sea ice (averages over September to November), March and September. The definition of winter and summer sea ice captures the months of minimum and maximum sea ice volume in the models which lags the annual cycle of insolation. The time series of seasonal or monthly averages have annual resolution and are then used to calculate the autocorrelation (with a lag of one year) and variance. With this approach we again focus on the effective relaxation time from year to year, and not the transient development of a perturbation within a year. In contrast to approach 1, the simulations involve stochastic terms that we add to the deterministic model equations. This involves choices on the place these terms are introduced in the equations (noise source), the magnitude of the noise (noise level) and its spectrum (color).

3. We analyse the trends of variance and autocorrelation in transient simulations of all models except E12 using a sliding window approach. As in the case of stationary simulations, all time series here are seasonal means, hence the time series have annual resolution. In particular, we analyse a simulation from MPI-ESM where $CO_2$ increases linearly until it has quadrupled after 2,000 years. This simulation has been performed and reported by Li et al. (2013). We perform similar simulations with E07, using different experiment lengths and 1,000 realisations per experiment. We also apply this method to the combined historical and RCP8.5 scenario simulations from CMIP5. Such transient simulations with continuously increasing $CO_2$ concentration more closely describe the ongoing change of the real climate system than the idealised, stationary experiments described before.

## 3   Results

### 3.1   Deterministic perturbation experiments

We begin by analysing how the response time τ of the model by Eisenman (2012) depends on the surface long-wave balance $L_m$ (Fig. 2a). In agreement with Eisenman (2012) and Moon and Wettlaufer (2011), this curve shows a characteristic double-peak with increasing $CO_2$ (decreasing $L_m$). The first peak occurs at $L_m \approx 1$ where the summer ice is lost and the ice-albedo feedback is substantially strengthened due to the exposed open water during a growing period during the year. The second peak occurs at the bifurcation point $L_m \approx 0.93$ where the winter sea ice vanishes. To this extent, the system is in agreement with dynamical-systems theory that predicts a slowing down as a result of increasing positive feedbacks.

To show that both peaks are indeed caused by the ice-albedo feedback, we perform additional experiments where this feedback is switched off. Following Eisenman (2012), we do this by setting the albedo difference between ice and water to 0. Appendix A explains the changes we make to the model equations in order to switch off certain mechanisms. Fig. 2b shows the relaxation time $\tau$ for the model without ice-albedo feedback but no other changes. Obviously, the range of $L_m$ over which a complete ice loss occurs is much larger due to the removal of a positive feedback. The most striking change in the evolution of $\tau$ is that the bifurcation as well as the double peak in the relaxation time have disappeared. The role of the ice-albedo feedback in E12 has also been analysed analytically by Moon and Wettlaufer (2011) who obtained the same result.

Another striking feature in Fig. 2a and Fig. 2b is the large regime of decreasing $\tau$ from preindustrial conditions up to shortly before the loss of summer sea ice. This decline results from the fact that the heat conduction through the ice becomes more efficient with decreasing thickness. This is important during freezing conditions when the heat from the ocean has to diffuse through the ice before it can be radiated away from the ice's surface. Therefore, thin ice grows faster than thick ice (Bitz and Roe, 2004), and the thinner the sea ice becomes, the more rapidly it can adjust to perturbations. Fig. 2c documents the validity of this interpretation: In addition to switching off the ice-albedo feedback, we also remove the growth-thickness feedback from the equations. To still obtain a stable system, the removed stabilising feedback is replaced by the negative Planck feedback that is also active in the ice-free season in the default model (see Appendix A). As a result, the relaxation time is constant in the regimes of perennial ice cover or open ocean.

Finally, a third alteration to the model reveals the reason for the increase of $\tau$ in the regime of seasonal sea ice (Fig. 2d). The difference to the version in Fig. 2b is that we halve the effective heat capacity of the ocean's mixed layer (e.g. representing a more shallow mixed layer). Obviously, this reduces the relaxation time for the ice-free system, because the model then simply consists of a well-mixed box of water whose heating or cooling rate is proportional to its mass. The warmer the climate becomes, the longer is the ice-free season, and the more does the system's effective timescale approach the timescale of an ice-free ocean. As this timescale is longer than the one of the thin sea ice, a "slowing down" occurs. Therefore, this increase in relaxation time is not related to any bifurcation approaching (there is none in Fig. 2b-d), or in fact to any positive feedback. This result has now also been obtained in a spatially explicit version of E12 (Wagner and Eisenman, in review).

The above result is also confirmed by the second column model, E07. Interestingly, for the loss of summer sea ice this model displays an evolution of $\tau$ that matches the results from E12 with fixed albedo (Fig. 2b): A regime of decreasing $\tau$ during the loss of summer sea ice is followed by a regime of slightly increasing $\tau$ after the complete summer sea-ice loss (Fig. 3a). For winter sea ice, in contrast, results with E07 match the evolution of E12 with interactive albedo, with a narrow peak in relaxation just before the loss of winter sea ice. This peak disappears when the albedo feedback is disabled in E07, while the response time for the loss of summer sea ice remeins largely unchanged (Fig. 3b). This demonstrates that the ice-albedo feedback is of secondary importance for the evolution and stability of summer sea ice in E07 (except at the very point of abrupt winter ice loss). It is important to note that both models roughly show the same evolution of the relaxation time (decreasing during summer ice loss, increasing thereafter), regardless of whether there is a bifurcation or any abrupt change approaching or not.

### 3.2 Stationary stochastic simulations

In natural systems, the relaxation time usually cannot be measured or calculated as directly as in models. However, one can hope to measure the system's response to natural external perturbations indirectly in form of its variance and autocorrelation. We therefore investigate in stochastic versions of the two column models whether these indicators reflect the changes in timescale. In each experiment, we introduce noise in one of three terms of the equations: the ocean heat flux into the mixed-layer (OHF), the long-wave energy balance (LW) and the short-wave insolation (SW). We distinguish small and large noise, as well as white and red noise. The way we introduce the stochastic terms to the equations and the details of the stochastic process are explained in Appendix A. It turns out that the specific choices of the noise terms hardly affect the results. When introducing small noise to the equations, the evolution of variance and autocorrelation closely follow the results we obtained from the perturbation experiments (Fig. 2), independent of the noise type. Even in case of large noise, the results are qualitatively similar as long as the noise is still small enough to not destroy the whole bifurcation structure of the system. Fig. 4 shows results for large red noise for all three noise sources as an example. The chosen decorrelation times of the red noise terms are of the same order of magnitude as in the real climate system (several days for atmospheric radiation and months for the ocean mixed layer). These time scales are still smaller than the typical response time of the ice-ocean system in the model (several years). In this regard, the model still sees the imposed noise as white, and the autocorrelation we find is determined by the system's time scale and not the time scale of the red noise. This explains the invariance of the results to the noise type.

We find the same robustness to noise source, level and colour in E07 (Fig. 5). Naturally, the variance of the summer ice area shows a distinct peak before the thickness approaches 0 because A becomes very sensitive to small perturbations in V (first row in Fig. 5). The peak occurs when fluctuations in A are least affected by the lower and upper limits, A=0 and A=1, and variance decreases thereafter because a larger fraction of the summer is already ice free, thus reducing the possible total variability. As sea ice disappears first in September, the variability peak occurs first in this month. The increased open-water formation efficiency before total ice loss has been reported in previous studies (Holland et al., 2006; Goosse et al., 2009) and is most evident during the gradual process of summer ice loss (around $CO_2 \approx 1.9$). The phenomenon is confined to a very narrow parameter regime for the abrupt winter ice loss around $CO_2 \approx 3.4$ because the rapid growth of new ice each winter tends to keep A close to 1 until shortly before total ice loss (Bathiany et al., in review).

The evolution of the volume's autocorrelation (last row in Fig. 5) closely follows the timescale obtained from the perturbation experiment. In principle, this is also true for the variance of ice volume fluctuations although it does not show a clear increase after summer ice loss (third row in Fig. 5). Interestingly, the autocorrelation of ice area (second row in Fig. 5) does not show the same evolution as the autocorrelation of ice volume in the regime of perennial ice. Such non-intuitive behaviour can occur as a result of the noise propagation through the nonlinear system and due to the permanent changes of feedbacks in different times of the year (Moon and Wettlaufer, 2013). Ice-area anomalies tend to have a shorter time scale than volume adjustments, especially due to the rapid growth of new ice that can quickly produce a large area increase with only small volume changes. Therefore, the autocorrelation of ice area is usually smaller than that of ice volume (it should be noted that the autocorrelation of the area fraction of winter sea ice has little practical relevance for most $CO_2$ levels because A is always very close to 1, as is reflected by its very small variance). As it is the slowest mode that dominates the relaxation time of the full system, the autocorrelation of ice volume corresponds well to the time scales we measured in the perturbation experiments (Fig. 2).

### 3.3 Transient stochastic simulations

We now analyse transient simulations with E07 and compare them to the most comprehensive model, MPI-ESM. We focus on the evolution of ice volume and its statistics. Each experiment starts from pre-industrial $CO_2$, which is then quadrupled over 2000 years. After approx. 1550 years, an abrupt loss of Arctic winter sea ice occurs in both models. We have tuned the sensitivity of E07 such that this event occurs at the same time in the models (Appendix B). As the description of the ice thickness distribution is similar in E07 and MPI-ESM, the abrupt winter sea-ice loss probably results from the same threshold mechanism. This is corroborated by the fact that the abrupt loss is reversible in MPI-ESM (Li et al., 2013).

We use red noise that is added to the ocean heat transport and that has a noise level which produces a similar variability in ice volume as MPI-ESM at individual grid cells. To obtain the evolution of variance and autocorrelation of the ice volume in both models we apply the 'early warnings' R package described in Dakos et al. (2012a), which performs an analysis often applied to transient time series (Lenton, 2011). The method consists in a running window of 300 years that slides from the beginning of the time series to the point just before the ice loss. In the case of summer sea ice, this final point is reached after 800 years, in the case of winter ice loss after 1550 years. As in the case of the stationary simulations, each time series consists of annually spaced seasonal means. Within the running window, fluctuations on long timescales are removed by smoothing the time series with a Gaussian kernel of interactive bandwidth and subtracting this smoothed version from the original time series. For the residuals, variance and autocorrelation are calculated within the window. As the window moves along the time series, we obtain an evolution of variance and autocorrelation (being shorter than the original time series by the window length).

The results with E07 are similar to the stationary experiments in the previous section (Fig. 6). As the simulations are much shorter than the stationary experiments, the results are much noisier. However, the decrease in variance (Fig. 6 d-f) and the decline and subsequent increase in autocorrelation of V (Fig. 6 g-i) are still clearly visible. As MPI-ESM is a spatially explicit model, one has to choose a specific region. We analysed six different single grid cells in the Arctic ocean and obtain a qualitatively similar evolution of statistics; Fig. 6 b, e, and h show results for a grid cell located at approx. 102 W and 86.5 N. Fig. 6 c, f, and i show results for the total ice volume north of 75 N. Thus, the behaviour at individual grid cells carries over to the regional scale. The results from MPI-ESM are also in good agreement with the results from E07 – the inclusion of spatial differences and processes like advection and mechanical redistribution of sea ice apparently has not changed the behaviour of sea ice variability. We therefore argue that E07 is an appropriate model to explain the behavior in MPI-ESM and it is probable that the same processes are behind the evolution of the statistics.

As Fig. 6 only presents a single realisation from both models the question arises how long a time series needs to be in order to observe significant trends. We therefore conducted four different experiments with E07 where the quadrupling of $CO_2$ occurs over 100, 200, 500 and 2000 years, respectively. For each experiment we perform 1,000 realisations and calculate the trends in variance and autocorrelation in each realisation. These trends are given as Kendall Tau values that express how monotonically a property changes. A time series with only positive (negative) changes from one point to the next has a Kendall Tau of 1 (-1), a time series with an equal number of increases and decreases has a Kendall Tau of 0. Fig. 7 shows the distribution of Kendall Tau values for the trends in variance and autocorrelation of winter sea ice. Sea ice loss occurs at slightly different times in the different realisations. Winter sea-ice loss typically occurs after 4/5 of the experiment length. In each realisation, the sliding window in which variance and autocorrelation are measured therefore stops 5 years before zero ice volume occurs for the first time in September (Fig. 7a,b) or March (Fig. 7c). Increasing the window length improves the results, but the window length is of course limited by the length of the time series. We therefore chose a constant relative window length of 3/20 of the experiment length.

The results somewhat depend on the details of this analysis and the system under consideration. However, Fig. 7 illustrates that several hundred to thousand years are required to obtain robust trends. While these results support our interpretation that the 2,000-year experiments in Fig. 6 are meaningful, simulations with more plausible scenarios cannot be expected to yield robust results. In general, variance is better constrained than autocorrelation (Ditlevsen and Johnson, 2010). Therefore, one can expect to see a decrease in variance of sea-ice volume but no consistent changes in autocorrelation in simulations where sea-ice loss occurs within less than 200 years, a typical experiment length for projections of anthropogenic climate change. This is confirmed when analysing the historical and RCP8.5 simulations from MPI-ESM and seven other comprehensive climate models (Fig. 8). While variance decreases in most models (especially those with a large pre-industrial variability), autocorrelation shows no convincing signal compared to Fig. 6. The decrease in variance occurs at individual grid cells and is thus likely to result from the increasing growth-thickness feedback discussed in Sect. 3.1. As we have shown in the previous sections, the trends in variance and autocorrelation that occur in sufficiently long simulations are not specific to the mechanism of ice loss. Fig. 7-8 illustrate yet another limitation to the applicability of early warning signals: Even if there was any information in these trends, it would be impossible to detect it in a single realisation with the current rate of global warming.

## 4 Conclusions

We have shown that the relaxation time of Arctic sea ice tends to decrease before summer ice loss and to increase before winter ice loss in three different models. In time series of sea ice volume these trends carry over to autocorrelation and, to some extent, variance. The decrease in response time during summer sea-ice loss is caused by the increased heat flux through the thinning ice; the increase during winter sea-ice loss is mainly caused by the long response time of the ocean which becomes more influential as the ice retreats. This characteristic relation between the sea ice state and its variability is robust in the two box models and also occurs in the available simulation with a comprehensive model. At first sight, our results may appear to be in conflict with the classical concept of slowing down. In principle however, the concept does apply to the case of sea ice loss: Just before the bifurcation at the point of winter sea ice loss occurs in E12 or E07, a sharp peak emerges (Fig. 2a, Fig. 3a). The peaks are more pronounced when the ice-albedo feedback is important like in E12, where τ even peaks during summer ice loss, and less pronounced before the winter ice loss in E07 which is mainly due to a threshold mechanism.

The practical problem is that these bifurcation-induced peaks occur in such a narrow parameter regime that it will be impossible to detect them before an abrupt change in transient time series. The general trends in transient time series will therefore be independent of the mechanism or even the existence of an abrupt change. In order to infer information on the system from its variability, these trends would need to be more specific to certain mechanisms. In models of low or intermediate complexity however, it may well be possible to investigate the mechanism of an abrupt change diagnostically by creating long stationary time series for carefully selected forcing conditions (Bathiany et al., 2013). It is therefore useful to see that our results are robust to the source of the noise, its spectrum and its magnitude. We have also shown that long simulations are necessary to obtain robust results, typically more than 1000 years (also see Ditlevsen and Johnson, 2010). This is also the reason why it will be difficult to see consistent trends in observations. Livina and Lenton (2013) found a recent increase in autocorrelation for summer ice area from satellite observations when corrected for the continental distribution, but no other clear signals due to the rather short record. The column models we applied indicate that ice thickness (or specific volume in the models), captures the system's relaxation time better than area fraction. Unfortunately, ice thickness is much more difficult to observe than sea ice coverage.

These results have implications for many other systems whose effective 'mass' changes over time, affecting the trend of their relaxation time. For instance, the effective heat capacity of the ocean can change with the mixed layer depth (Boulton and Lenton, 2015). Moreover, the relaxation timescale may depend on the direction of perturbations, just like sea ice melting and freezing is determined by different processes. Another example for such asymmetry is vegetation dynamics (Bathiany et al., 2012): While vegetation can die back or burn within days or months, its regrowth can take many decades. Such restrictions and the fact that the statistics of sea ice in the models are closely linked to its mean state may make the prospect of 'early warnings' for accelerated sea ice loss appear rather limited. However, this also provides an opportunity: A strict relation between the mean state of sea ice and its variability would allow an estimate of the system's total variability from relatively short observational time series. In this regard, the state of the system and its trend can provide an early warning for potential extreme events as their magnitude and longevity depends on the mean state. This knowledge could be important for ecosystems and economical activities in the high latitude oceans.

**Appendix A: Description of Eisenman (2012) model**

Here, we describe the model by Eisenman (2012), denoted E12 in the main text, and the changes made to separate different effects.

The dynamic equation of the model is

$$\frac{dE}{dt} = A - BT + F_B \qquad (A1)$$

with t for time and E for enthalpy. In the presence of sea ice, E is negative and proportional to the ice's thickness, while during ice-free conditions, E is positive and proportional to the mixed-layer temperature.

Term A in Eq. A1 describes the temperature-independent terms of the radiative balance

$$A = \left(1 + \Delta_\alpha \tanh\left(\frac{E}{h_\alpha}\right)\right)(1 - S_a \cos 2\pi t) - L_m - L_a \cos 2\pi(t - \Phi) \qquad (A2)$$

with $L_m$ is the annual mean long-wave radiation balance at the surface, the control parameter we vary in our experiments to represent a change in atmospheric $CO_2$.

T represents the surface temperature of the ice-ocean system and is calculated from

$$T = \begin{cases} E, & E \geq 0 \quad \text{[open ocean]} \\ 0, & E < 0, A > 0 \text{ [melting surface]} \\ \frac{A}{B}\left(1 - \frac{\zeta}{E}\right)^{-1}, & E < 0, A < 0 \text{ [frozen surface]} \end{cases} \qquad (A3)$$

These equations correspond to Eq. 9-11 in Eisenman (2012) with the exception that we have omitted the tilde above all variables that denotes them as non-dimensional variables. All parameter values are listed in his Tab. 1. For a derivation of these equations and an explanation of all parameters see Eisenman (2012) and Eisenman and Wettlaufer (2009).

To switch off the ice-albedo feedback, we set Δα to zero. To switch off the growth-thickness feedback in addition, the temperature equation is replaced by

$$T = \begin{cases} E, & E \geq 0 \text{ [open ocean]} \\ 3E, & E < 0 \text{ [sea ice]} \end{cases} \qquad (A4)$$

This way, the stabilising growth-thickness feedback is replaced by the stabilising Planck feedback, the same that also operates under ice-free conditions. The factor 3 in the presence of sea ice is arbitrary and was introduced merely to distinguish the regime with and without sea ice in Fig. 2c.

Alternatively, to reduce the heat capacity of the mixed layer by a factor 2 we exchange the temperature equation by

$$T = \begin{cases} 2E, & E \geq 0 \quad \text{[open ocean]} \\ 0, & E < 0, A > 0 \text{ [melting surface]} \\ \frac{A}{B}(1 - \frac{\zeta}{E})^{-1}, & E < 0, A < 0 \text{ [frozen surface]} \end{cases} \quad (A5)$$

As the equations are dimensionless, the mixed-layer heat capacity C does not explicitly appear in the model equations. In case of open water, E incorporates the inverse of C, which is why halving C corresponds to doubling E in the open water case of the above equation (see Eisenman (2012) for details on the model derivation).

In the deterministic numerical perturbation experiments, we perturb E by 0.005 and measure the decay rate over two years. The decay rate is thus determined from three points (start time, end of year one, and end of year two). Using more years leads to the same results but fails in cases when the system is very stable because the anomaly then becomes too small to be measured after only a few years. For the stationary simulations with noise in the ocean heat flux ($\sigma_{OHF}$), we added a Gaussian white noise term to Eq. A1. To introduce noise to the radiative fluxes, we added the noise term on the radiative balance A (Eq. A2) to perturb the long-wave balance ($\sigma_{LW}$), or on S ($S = 1 - S_a \cos 2\pi t$) in Eq. (A2) to perturb the short-wave balance ($\sigma_{SW}$). In the case of small noise, we chose the noise level in a way that the total variance of E is in the order of $10^{-9}$, i.e. much smaller that the amplitude of an annual cycle. In the case of large noise, we adjust the noise level such that the system's stochastic variability is roughly one order of magnitude smaller than the amplitude of the annual cycle, in similarity to the situation in the real world. In case of red noise, we model the external perturbations as an autoregressive process of order one (AR(1) process) with a decorrelation time of 180 days in case of mixed layer energy and 10 days for atmospheric radiation.

**Appendix B: Description of Eisenman (2007) model**

In contrast to the model by Eisenman (2012), which consists of only one dynamic equation, four variables are explicitly modelled by ordinary differential equations: the ice area coverage A, the ice volume V, the surface temperature of the ice $T_i$, and the temperature of the mixed layer, $T_{ml}$. Therefore, the ice has only a single thickness h=V/A. The evolution of ice area is described by a parameterisation based on Hibler (1979). As in E12, the temperature profile within the ice is assumed to be linear. For the model equations including their derivation see Eisenman (2007). Due to the fact that many processes have been intentionally neglected, the original model is rather insensitive to $CO_2$. To obtain a similar sensitivity than with the comprehensive model MPI-ESM we have added an additional flux of 16 $Wm^{-2}$ per $CO_2$ doubling to the downwelling long-wave radiation at the surface (eq. 30 in Eisenman (2007)).

Due to the model's four state variables, it had to be decided how to perturb the system in the numerical perturbation experiments. In principle, a system responds differently depending on which state variable is perturbed. While the water's large specific heat capacity and latent heat of fusion determine the long-term slow response of the system, perturbations of the radiative fluxes decay very quickly. Our numerical perturbation experiment for E07 consists in a perturbation of $T_{ml}$ by +0.2 K. For the determination of the relaxation time via regression, we use years 2 and 3 after the perturbation is applied, ensuring that anomalies of the fast modes have already decayed after the first year. In the stochastic experiments, we introduce the stochastic terms in the same way as in the case of E12 (see Appendix A).

**Acknowledgements**

This work was carried out under the programme of the Netherlands Earth System Science Centre (NESSC), financially supported by the Ministry of Education, Culture and Science (OCW). We also acknowledge the World Climate Research Programme's Working Group on Coupled Modelling, which is responsible for CMIP, and we thank the climate modeling groups for producing and making available their model output. We are grateful to Vasilis Dakos for help with applying his early warnings R package and to Chao Li for making available the MPI-ESM model output. We are also indebted to Till Wagner and Ian Eisenman for their valuable comments and their very amiable and cooperative spirit.

**Figures**

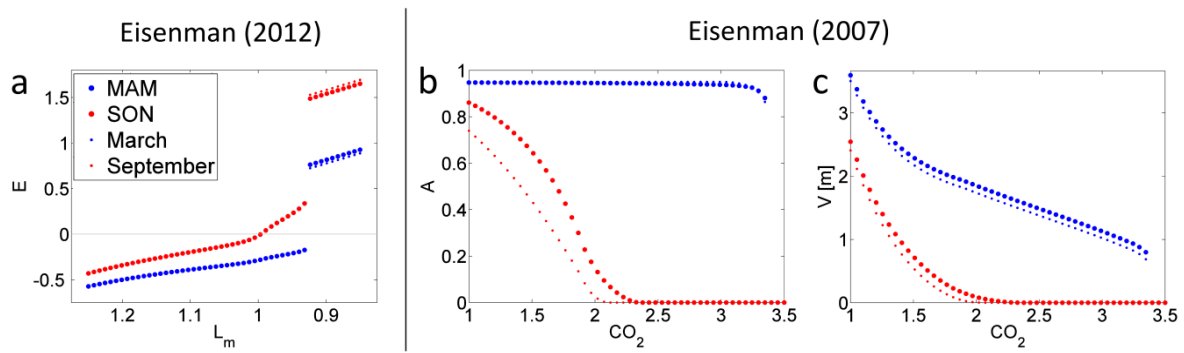

Figure 1. Response of the two column models to warming. a) Enthalpy versus surface longwave balance $L_m$ in E12. The horizontal line demarcates between positive E (open water) and negative E (ice covered ocean). b) Ice-area fraction and c) ice volume (given as an equivalent thickness) versus $CO_2$ (given as multiples of the pre-industrial value) in E07. Each dot represents a time mean over the season indicated in the legend of subfigure (a).

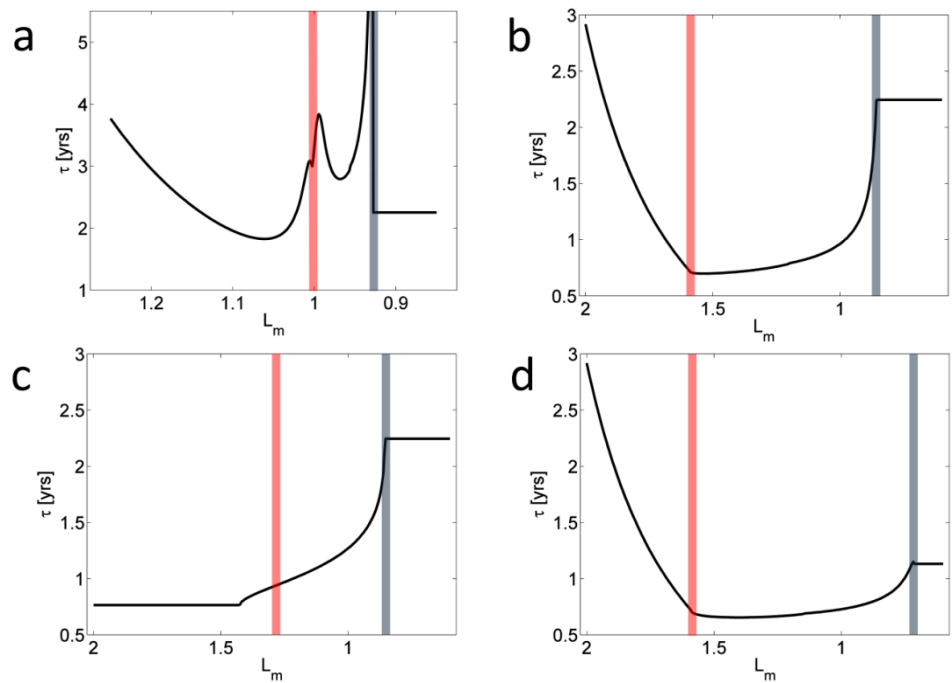

Figure 2. Relaxation time scale in Eisenman's (2012) box model for different combinations of mechanisms. a) Original model, b) original model but with disabled ice-albedo feedback; c) like (b) but without growth-thickness feedback; d) like (b) but with only half the default ocean heat capacity. The vertical shaded lines indicate the values of $L_m$ where summer ice (red) and winter ice (blue) disappear.

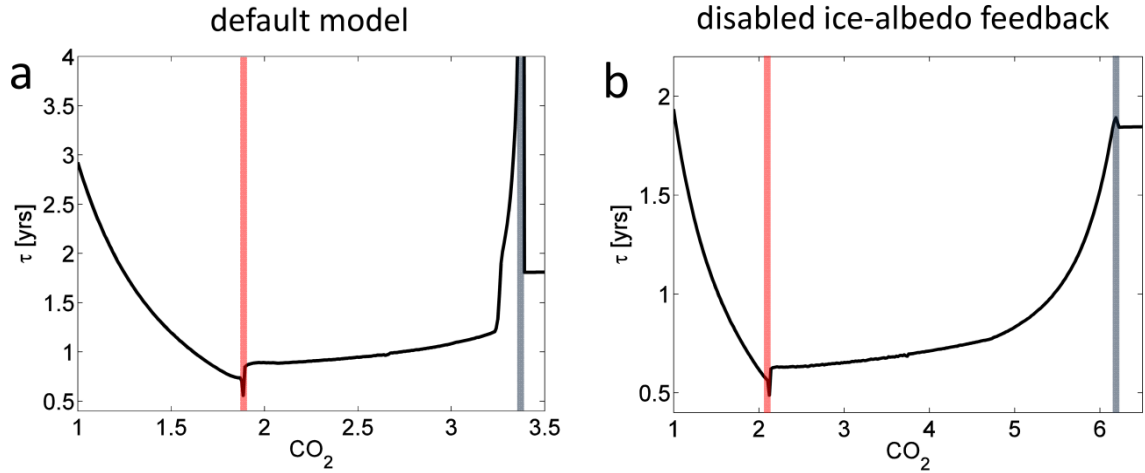

Figure 3. Relaxation time scale in Eisenman's (2007) box model for a) the original model, b) with disabled ice-albedo feedback. The vertical shaded lines indicate the values of $CO_2$ where summer ice (red) and winter ice (blue) disappear.

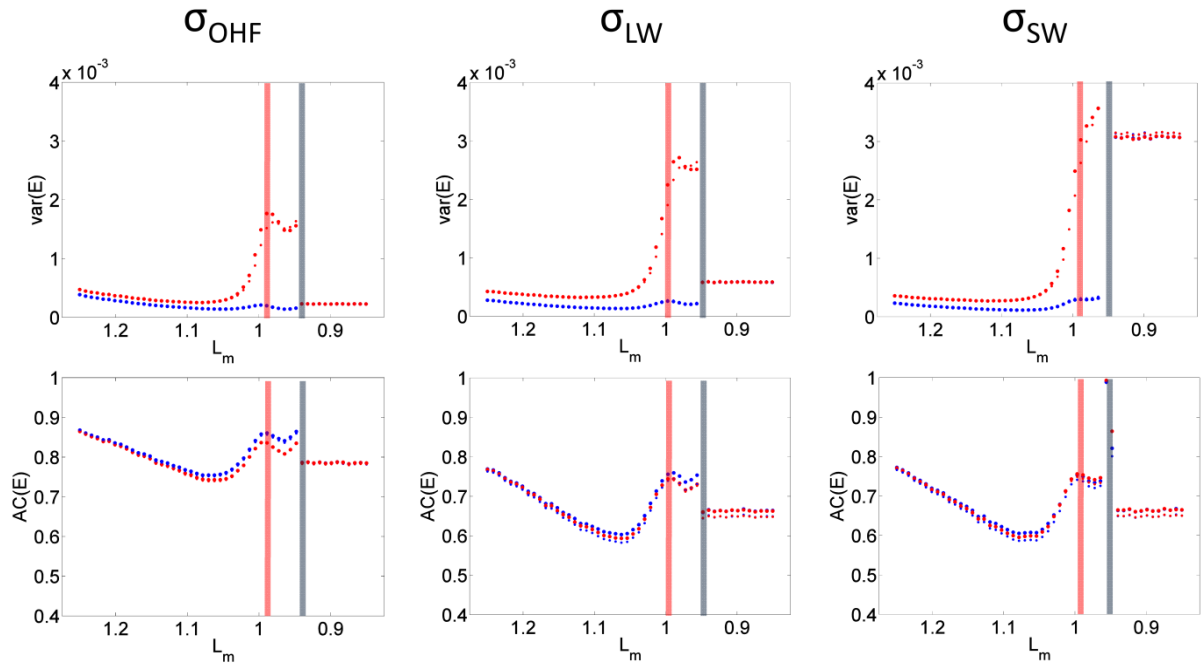

Figure 4. Variance (var) and autocorrelation (AC) of state E as a function of long-wave balance $L_m$ in the model E12 with large red noise. In each column, the noise term has been introduced to one of three different terms, namely the ocean heat flux (OHF), long-wave radiative balance (LW) and short-wave radiative balance (SW). The vertical shaded lines indicate the values of $L_m$ where summer ice (red) and winter ice (blue) disappear.

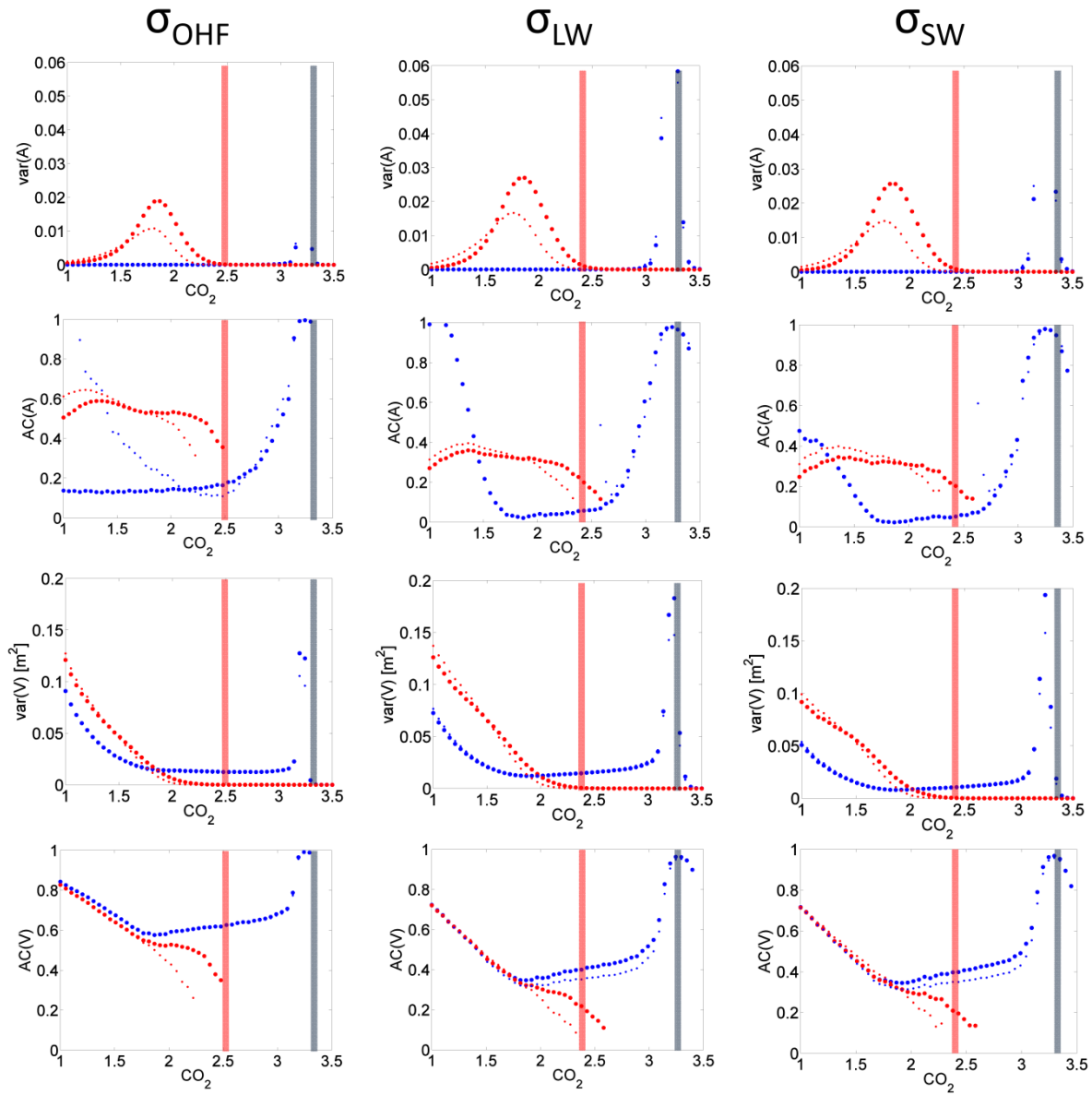

Figure 5. Variance (var) and autocorrelation (AC) of ice area fraction A and ice volume V as a function of $CO_2$ in the model E07 with large red noise. In each column, the noise term has been introduced to one of three different terms, namely the ocean heat flux (OHF), long-wave radiative balance (LW) and short-wave radiative balance (SW). The vertical shaded lines indicate the values of $CO_2$ where summer ice (red) and winter ice (blue) disappear.

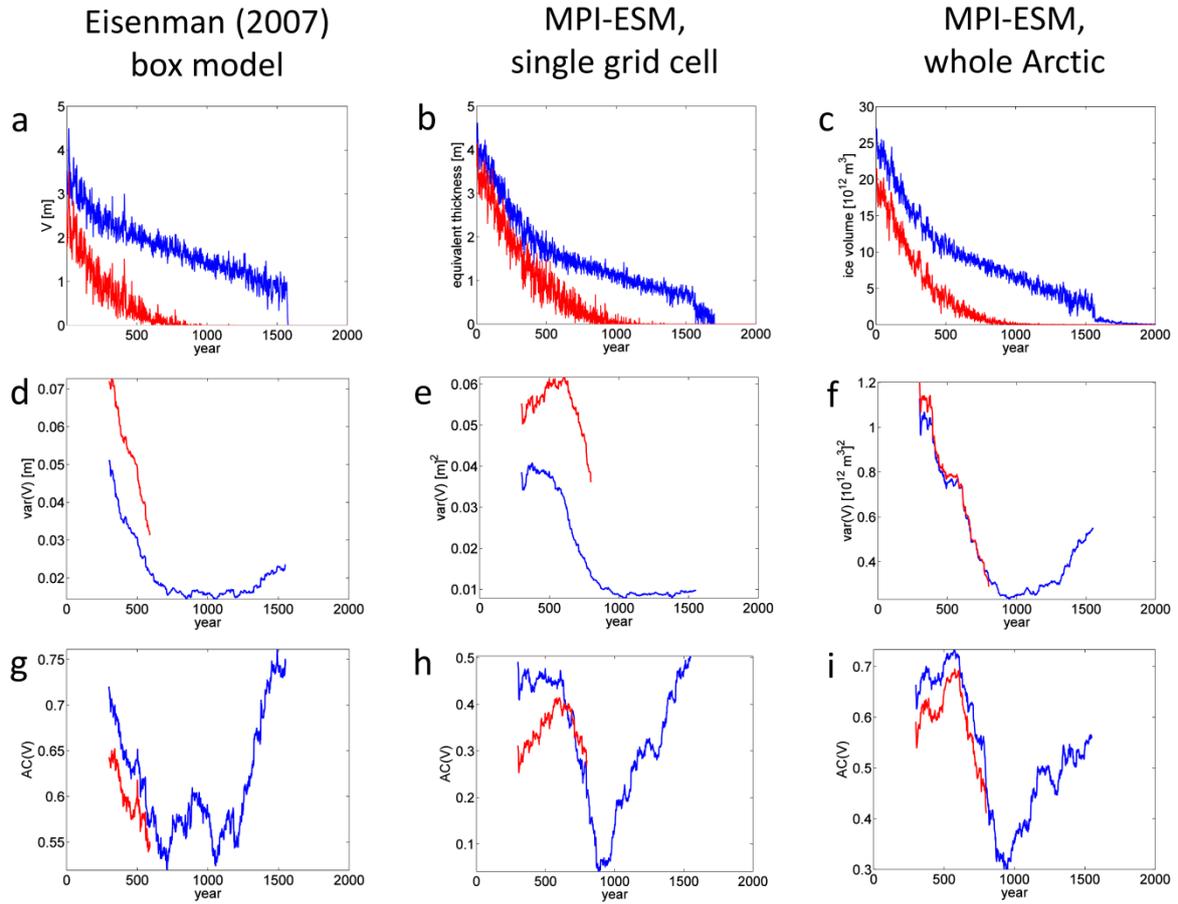

Figure 6. Time series from transient simulations with the box model by Eisenman (2007) (left column) and MPI-ESM (middle and right column). (a)-(c) evolution of ice volume; (d)-(f) variance, and (g)-(i) autocorrelation of this volume as obtained from a sliding window approach. Winter sea ice is shown in blue, summer sea ice in red. The single grid cell in MPI-ESM (second column) is located at 86.5 N and 102 W.

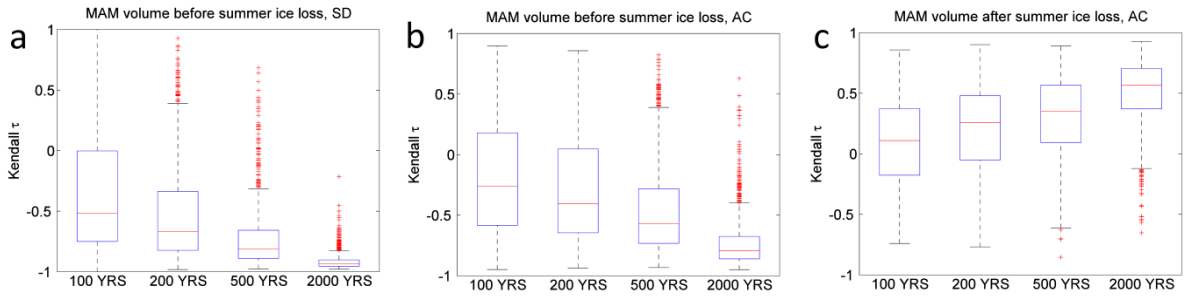

Figure 7. Statistics of Kendall's Tau for standard deviation (SD) and autocorrelation (AC) changes in ensemble simulations with the E07 model. Each figure shows results for MAM timeseries of sea-ice volume. The number of years refer to the total length of an experiment until $CO_2$ has quadrupled. a) standard deviation trends before summer sea-ice loss b) autocorrelation trends before summer sea-ice loss (perennial ice regime), c) autocorrelation trends in the period between summer sea-ice loss and winter sea-ice loss (seasonal ice regime). On each box, the central mark is the median, the edges of the box are the 25th and 75th percentiles, the whiskers extend to the most extreme data points not considered outliers, and outliers are plotted individually.

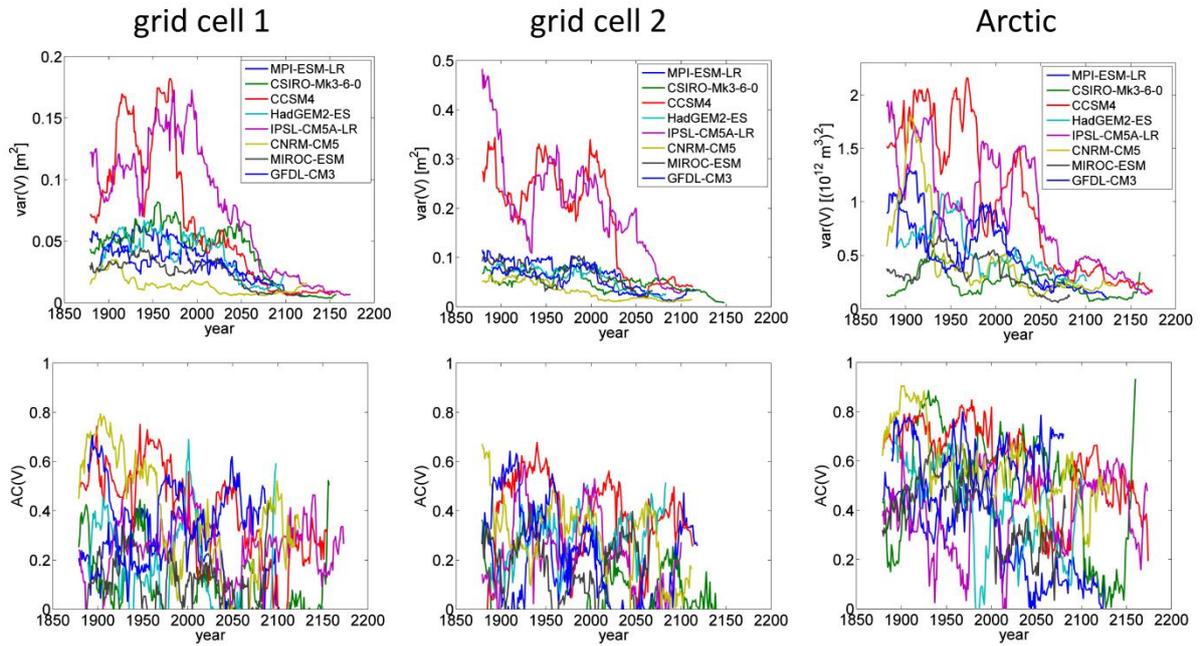

Figure 8. Evolution of variance (var) and autocorrelation (AC) of sea-ice volume (V) in eight comprehensive climate models. The time series of V (not shown) are the combined historical and RCP8.5 simulation, the window length for the calculation of var and AC is 30 years. Grid cell 1 is located at 102 W and 86.5 N, grid cell 2 at 180 W and 74.5 N.